\documentclass[prb,twocolumn,superscriptaddress,showkeys,floatfix]{revtex4-2}

\usepackage{graphicx}

\usepackage{bm}
\usepackage{xcolor}
\PassOptionsToPackage{hyphens}{url}\usepackage{hyperref}
\usepackage{url}
\usepackage{bookmark}
\usepackage{amsmath}
\usepackage[normalem]{ulem}

\newcommand{\be}{\begin{equation}}
\newcommand{\ee}{\end{equation}}

\def\bea{\begin{eqnarray}}
\def\eea{\end{eqnarray}}

\begin{document}

\title{Raman response in superconducting multiorbital systems with application to nickelates}

\author{Matías Bejas}
\affiliation{Facultad de Ciencias Exactas, Ingeniería y Agrimensura and Instituto de Física Rosario (UNR-CONICET),
Avenida Pellegrini 250, 2000 Rosario, Argentina}
\author{Jun Zhan}
\affiliation{Beijing National Laboratory for Condensed Matter Physics and Institute of Physics, Chinese Academy of Sciences, Beijing 100190, China}
\affiliation{School of Physical Sciences, University of Chinese Academy of Sciences, Beijing 100049, China}

\author{Xianxin Wu}
\affiliation{Institute for Theoretical Physics, Chinese Academy of Sciences, Beijing 100190, China}
\author{Andreas P. Schnyder}
\affiliation{Max Planck Institute for Solid State Research, Heisenbergstrasse 1, D-70569 Stuttgart, Germany}
\author{Andrés Greco}
\affiliation{Facultad de Ciencias Exactas, Ingeniería y Agrimensura and Instituto de Física Rosario (UNR-CONICET),
Avenida Pellegrini 250, 2000 Rosario, Argentina}

\date{\today}

 \begin{abstract}

The recent discovery of high-$T_c$ superconductivity in pressurized and thin film nickelates is nowadays one of the most relevant and active topics in solid-state physics. 
The origin of superconductivity together with the relevance of multiorbital physics
are highly discussed issues in this field.
Knowledge of the size of the gap and its symmetry is of fundamental interest
to uncover the superconducting mechanism at play in the nickelates. 
Electronic Raman scattering is a powerful tool to investigate the main characteristics of the gap. 
Here, we investigate  the Raman response in the superconducting phase for three different models:
Two-orbital models, including $d_{x^2-y^2}$ and $d_{z^2}$
orbitals, with one and two layers; as well as
a bilayer model with the $d_{x^2-y^2}$ orbital
as the only active one.
For each of these models, we consider
different pairing symmetries and determine their
characteristic fingerprints in the Raman response. 
For the two-orbital models, we perform full multiorbital calculations including interorbital and intraorbital scattering, and compare the results with those obtained using the additive Raman response where each band is considered separately. Our results should be useful for discussing the minimal model for superconductivity and its pairing symmetry in nickelates. The obtained results and discussions, as well as the presented formalism,  are also of general interest for other multiorbital systems. 
 
\end{abstract}

\maketitle

\section{Introduction} \label{sec:Intro}

Raman scattering is a well-known valuable experimental tool 
to investigate superconductors. From these experiments, in principle, we can extract information about the size of the superconducting gap and its symmetry\cite{devereaux07}. Although the theory behind Raman experiments in superconductors is old, the topic is not fully settled. The reason is easy to understand, and  is related to the fact that the power of theory-experiment feedback requires high-$T_c$ superconductors, which were discovered in the last years. The high value of the superconducting critical temperature $T_c$ implies a large superconducting gap, and that means that the pair breaking features across $T_c$ appear in an energy window accessible for Raman experiments. Clearly, this situation does not occur in conventional or low-$T_c$ superconductors. 

The discovery of high-$T_c$ cuprates superconductors\cite{keimer15,timusk99} boosts Raman studies 
in the superconducting phase where the size and symmetry of the superconducting gap was discussed extensively (see Ref. [\onlinecite{devereaux07}] for a review). 
While Raman in single-layer cuprates can be discussed in the context of a one-band model\cite{andreasthesis}, the bilayer cuprates require to deal with a multiband model\cite{devereaux96}, which is also required for multiorbital systems as iron pnictides\cite{boyd09,sauer82}.
One common approximation for multiband systems assumes that Raman scattering in the superconducting state can be discussed computing additively the Raman response from each band separately\cite{devereaux96,boyd09,sauer82}. In this approximation the Raman vertices are obtained in the effective mass approximation as second derivatives of each band.

Superconductivity in nickelates was first achieved in “infinite-layer" nickelates (Sr,Nd)NiO$_2$ thin films on substrates with a transition temperature ($T_c$) around 5-15 K ~\cite{Li2019,Osada2020,Pan2022,wang2022,ding_critical_2023}.
Similar to $d^9$ cuprates, the low-energy states are dominated by the Ni $d_{x^2-y^2}$ orbital but the contribution of Nd $s$ orbitals introduce a self-doping in the Ni $d_{x^2-y^2}$ orbital~\cite{sakakibara20,Wu2020,GMZhang2020}.
The $T_c$ of the infinite-layer nickelates can be enhanced to be over 30 K with external pressure~\cite{wang2022} and the pairing symmetry remains unclear~\cite{worm24,gu2020,harvey2022,chow2023,grissonnanche24,cheng24}.
Remarkably, high-$T_c$ superconductivity has recently been observed in the bilayer nickelates La$_3$Ni$_2$O$_7$ under moderate pressure, achieving $T_c$ values of approximately 80 K \cite{sun23,hou23}, thereby generating tremendous research activity.
The field was even more boosted by the recent discovery of superconductivity at ambient pressure in thin films\cite{ko25,zhou25} of La$_3$Ni$_2$O$_7$.
Different from cuprates it seems that a multiorbital character, where the active orbitals are $d_{x^2-y^2}$ and $d_{z^2}$, should be considered in these materials. See Ref. \cite{Wang25RR} and references therein for a complete review about the very recent progress in the field. After observing superconductivity at ambient pressure it is likely that Raman experiments can be performed in such materials, giving important information about the size and symmetry of the superconducting gap. 
The superconducting gap value and its symmetry is under debate and even showing contradicting results\cite{guo25,cao25}; see also Ref. \cite{Wang25RR} for discussions.
Theoretical studies on bulk and thin films nickelates showed tendencies to $d$ and $s^{\pm}$ pairing symmetry\cite{gu23,zhang24,zhang23R,liu23t,jiang25R,sakakibara24R,zhan25R,shao25R}. Other proposals suggest that the minimal model is a bilayer one-orbital model, where only the $d_{x^2-y^2}$ plays the most important role\cite{lu24,qu24,oh24t,bejas25}. 

In this paper we study the Raman response for different and basic multiorbital models, distinct superconducting gaps as $d$, $s^{\pm}$, and $s$, and Raman symmetries.
For the $d$-wave case we analyze two different cases, the interorbital $d$-wave pairing\cite{zhan25R}, and the intraorbital $d$-wave pairing on nearest-neighbors bonds\cite{maier25}.
The goal of our paper is to present qualitative features, and in addition to show that our calculation can be easily extended to others more complicated models inspired on different experimental and/or theoretical situations. In principle, our results should shed light into the size of the gap, its symmetry, and the minimal model accurate for discussing superconductivity.  
Although our paper was initially motivated by the recent discovery of superconductivity in nickelates, we present our results in a general context to facilitate its use in other cases where it is necessary to deal with multiband and/or multiorbital systems.

The paper is organized as follows. 
In Sec. \ref{sec:formalism} we describe the two formalisms used in the calculation; the full multiorbital (MO) calculation, and the additive response [(IB) for isolated bands] approximation where the Raman response is computed additively from the contribution of each band separately.
In Sec. \ref{sec:bi1orb} we discuss the bilayer one-orbital model.
In Sec. \ref{sec:single2orb} and Sec. \ref{sec:bi2orb} we present the results for the single-layer two-orbital model and the bilayer two-orbital model, respectively. Conclusions are discussed in Sec. \ref{sec:conc}.

\section{Formalism}\label{sec:formalism}

\subsection{ Full multiorbital Raman response} \label{sec:fullmulti}

In this section, for clarity and without loosing generality,  we discuss a single-layer two-orbital model,
calling these two orbitals $x$ and $z$.
This is a minimal multiorbital case and, as we will see,  it is easily extended to more orbitals and/or more layers.  

The proposed model is defined by the Hamiltonian $H_{0} = \sum_{\bf k} \psi_{\bf k}^\dag H_{\bf k} \psi_{\bf k}$, with
\begin{align}
  H_{\bf k} &=  
  \begin{bmatrix}
  t_{xx}({\bf k}) & t_{xz}({\bf k}) \\
  t_{zx}({\bf k}) & t_{zz}({\bf k})
  \end{bmatrix},
  \label{eq:Hk}
\end{align}
\noindent where $\psi_{\bf k}^\dag = (x_{\bf k}^\dag, z_{\bf k}^\dag)$, and $x_{\bf k}^\dag$ ($z_{\bf k}^\dag$) is the creation operator for electrons in the orbital $x$ ($z$).
Thus, the Hamiltonian $H_{\bf k}$ is a $2\times2$ matrix in the orbital basis.

In Eq. (\ref{eq:Hk}) 
$t_{xx}$ is the hopping between $x$ orbitals, 
$t_{zz}$ is the hopping between $z$ orbitals, and
$t_{xz}$ the hopping between the $x$ and $z$ orbitals ($t_{zx}$ is the complex conjugated of $t_{xz}$).
${\bf k}$ is the momentum in the two-dimensional (2D) lattice, where on each site the $x$ and $z$ orbitals are located.

The Hamiltonian in the band basis, $H_{\bf k}^B = U^\dag H_{\bf k} U$, is diagonal
\begin{align}
  H_{\bf k}^B &=
  \begin{bmatrix}
  e_1({\bf k}) & 0 \\
  0 & e_2({\bf k})
  \end{bmatrix},
  \label{eq:HkB}
\end{align}
\noindent where $e_1({\bf k})$ and $e_2({\bf k})$ are the bands, and  $U$  is the rotation matrix to go from the orbital basis to the band basis. The momentum dependence in $U$ was omitted for simplicity. 

The Raman vertices in the orbital basis are $2\times 2$ matrices given by
\begin{align}
R_{\bf k}^\gamma &= \left(\frac{\partial^2}{\partial k_x^2} \pm \frac{\partial^2}{\partial k_y^2}\right)H_{\bf k} /, ,
\end{align}
\noindent where $+$ ($-$) defines the symmetry $\gamma=A_{1g}$ ($\gamma=B_{1g}$) vertex, and 
\begin{align}
R_{\bf k}^\gamma &= 2 \left(\frac{\partial^2}{\partial k_x\partial k_y}\right)H_{\bf k} 
\end{align}
\noindent for $\gamma=B_{2g}$. 

The Raman vertex in the band basis, $R_B^\gamma = U^\dag R^\gamma_{\bf k} U$, reads 
\begin{align}
  R_B^\gamma &=
  \begin{bmatrix}
  R_{11}^\gamma & R_{12}^\gamma \\
  R_{21}^\gamma & R_{22}^\gamma
  \end{bmatrix} \, ,
  \label{eq:RkB}
\end{align}
\noindent where $R_{11}^\gamma$ and $R_{22}^\gamma$ correspond to intraband transitions, while $R_{12}^\gamma$ and $R_{21}^\gamma$ to interband. 

For the calculation in the superconducting state we introduce the Nambu spinor in the band basis
\begin{align}
{\Psi_{\bf k}^B}^\dagger=(c^\dagger_{1,{\bf k}\uparrow}, c^\dagger_{2,{\bf k}\uparrow}, c_{1,{\bf -k}\downarrow}, c_{2,{\bf -k}\downarrow}), 
\end{align}
\noindent where $c^\dagger_{\alpha,{\bf k}\sigma}$ is the electron creation operator in the band $\alpha$, momentum ${\bf k}$, and spin $\sigma$.
The pairing gap is given in the orbital basis and has the general matrix form:
\begin{align}
  \Delta &=
  \begin{bmatrix}
  \Delta_{xx} & \Delta_{xz} \\
  \Delta_{zx} & \Delta_{zz}
  \end{bmatrix}.
  \label{eq:gaporb}
\end{align}

In the band basis the pairing gap is given by $\Delta^B = U^\dag \Delta U$. Here, interband pairings are neglected with a further assumption that the weak Cooper pairing takes place only between electrons on the same band. Thus, we only consider the diagonal elements of the matrix $\Delta^B$ (only minor differences appear if the full $\Delta^B$ is used). If only interband gaps are considered, pair breaking peaks do not appear in the low-energy region below the onset of interband transitions.  Instead, a gap opens in the interband region and a weak redistribution of spectral weight occurs at high energy in the Raman response. However, this redistribution of spectral weight is not expected to be observed in experiments since at high energy the continuum from interband transitions is typically broad and contains many different contributions. In other words, intraband gaps are associated with low-energy intraband transitions and interband gaps with high-energy interband transitions. It is important to mention that neglecting interband gaps does not mean to neglect the mixing of the orbital gaps, this mixing is also contained in the intraband gaps.

In the Nambu notation the inverse of the Green's function reads  
\begin{align}
  G^{-1}({\bf k}, i\nu_n)&= 
i\nu_n \mathrm{I} - 
		\tau_3
		\otimes
        H_{\bf k}^B
        -
        \tau_1
		\otimes
		\Delta^B
 \nonumber \\
  &=\begin{bmatrix}
  i \nu_n-e_1& 0 & -\Delta_1 &0\\
     0 & i \nu_n-e_2 & 0 &-\Delta_2\\
  -\Delta_1& 0& i \nu_n+e_1& 0 \\
 0& -\Delta_2& 0& i \nu_n+e_2
 \end{bmatrix} \, ,
  \label{eq:Green}
\end{align}
\noindent where $\mathrm{I}$ is the $4\times 4$ identity matrix, $\tau_1$ and $\tau_3$ are the Pauli matrices, and $\Delta_1$ and $\Delta_2$ are the intraband superconducting gaps, and in general they depend on ${\bf k}$. $\nu_n$ is a fermionic Matsubara frequency. 

The Raman response $\chi^\gamma(i\omega_n)$ is calculated as the imaginary part of
\begin{align}
 \chi^\gamma(i\omega_n) &= \nonumber \\
 \sum_{{\bf k},i\nu_n} &
 \mathrm{Tr}\left\{ [\tau_3 \otimes R_{B}^\gamma] G(i\nu_n) [\tau_3 \otimes R_{B}^\gamma] G(i\nu_n+i\omega_n) \right\}
\label{eq:fullRaman}
\end{align}
\noindent after the analytical continuation $i\omega_n = \omega + i\Gamma$. The value of $\Gamma$ is positive and, in principle, infinitesimally small.
We chose $\Gamma/t = 5 \times 10^{-4}$ to better show the different peaks in the Raman response.
$\omega_n$ is a bosonic Matsubara frequency.
We omitted the momentum ${\bf k}$ dependence in the Raman vertices and the Green's functions for simplicity.

\subsection{Additive Raman response approximation}\label{sec:IB}
As mentioned in Sec. \ref{sec:Intro}, the Raman scattering in superconductors can be also studied using the IB approximation that considers the addition of the Raman responses from each band $e_\alpha({\bf k})$ separately\cite{devereaux96,boyd09,sauer82}.
Thus, in the IB approximation the Raman response $\chi_{\rm IB}^{\gamma}$ is given by
\begin{align}
\chi_{\rm IB}^{\gamma}(i\omega_n) = \sum_{\alpha} \chi_{\gamma \gamma, \alpha}(i\omega_n) \, ,
\label{Rasauer1}
\end{align}
\noindent where $\alpha$ runs over the number of bands, and $\chi_{\gamma \gamma, \alpha}$ is the bare Raman susceptibility calculated for a given band $\alpha$ as 
\begin{align}
 \chi_{\gamma \gamma,\alpha}(i\omega_n) &= \nonumber \\
 \sum_{{\bf k},i\nu_n}&
 \gamma_\alpha^2 \, \frac{\Delta^2_\alpha}{E^2_\alpha}
 \tanh \frac{E_\alpha}{2T}
 \left[ \frac{1}{i\omega_n + 2E_\alpha} - 
        \frac{1}{i\omega_n - 2E_\alpha}\right] \, ,
\label{eq:fullRaman_n}
\end{align}
\noindent where the momentum ${\bf k}$ dependence was omitted for clarity, $E^2_\alpha({\bf k}) = e^2_\alpha({\bf k}) + \Delta^2_\alpha({\bf k})$, $\Delta_\alpha({\bf k})$ is the gap for the band $\alpha$, and $T$ is the temperature. Note that in the IB approximation only intraband gaps are necessary.
The Raman vertices are
\begin{align}
\gamma_\alpha({\bf k}) &= \left(\frac{\partial^2}{\partial k_x^2} \pm \frac{\partial^2}{\partial k_y^2}\right)e_\alpha({\bf k}) \, ,
\end{align}
\noindent where $+$ ($-$) defines the $\gamma=A_{1g}$ ($\gamma=B_{1g}$) vertex, and 
\begin{align}
\gamma_\alpha({\bf k}) &= 2 \left(\frac{\partial^2}{\partial k_x\partial k_y}\right) e_\alpha({\bf k})
\end{align}
\noindent for $\gamma=B_{2g}$.

Equation (\ref{Rasauer1}) is considered only for $\gamma=B_{1g}$ and $\gamma=B_{2g}$.
For the $A_{1g}$ symmetry the Raman response is\cite{sauer82}
\begin{align}
\chi_{\rm IB,scr.}^{\gamma}(i\omega_n) = \chi_{\rm IB}^{\gamma}(i\omega_n) -
 \frac{\left[ \sum_\alpha \chi_{\gamma 1,\alpha}(i\omega_n) \right]^2}{\sum_\alpha \chi_{11,\alpha}(i\omega_n)} \, .
\label{Rasauer}
 \end{align}

The second term on the right-hand side of Eq.(\ref{Rasauer}) takes into account the Coulomb screening.
We use the notation where $\chi_{{\gamma\gamma},\alpha}$ means that the form factor in Eq. (\ref{eq:fullRaman_n}) is $[\gamma_\alpha({\bf k})]^2$, in $\chi_{{\gamma1},\alpha}$ is $\gamma_\alpha({\bf k})$, and in $\chi_{11,\alpha}$ is $1$.
The first term on the right-hand side of Eq.(\ref{Rasauer}) is just the free $A_{1g}$ response. 

It is instructive to make calculations using the IB approximation and compare the results with the obtained ones using the MO approach, because the additive approximation is frequently done and it might be useful for obtaining a first and qualitative view of the Raman response.

\section{Bilayer one-orbital model}\label{sec:bi1orb}

The model consists on two square lattice planes with hopping $t$ between nearest-neighbors sites on each plane and a hopping $t_\perp$ between nearest-neighbors sites on different planes. 
This model was recently proposed for bilayer nickelates in the context of the strongly correlated $t$-$J_\parallel$-$J_\perp$ model\cite{lu24,qu24,oh24t,bejas25}, where $J_\parallel$ and $J_\perp$ are the in-plane and out-of-plane magnetic exchanges, respectively. In this model, and similar to cuprates,  only the $d_{x^2-y^2}$ orbital plays an active role, where the hopping between planes is small\cite{luo23}.
In addition, this model may represent, in a first approximation, a situation for discussing infinite-layer nickelates where only the $d_{x^2-y^2}$ orbital was proposed to play the main role\cite{worm24}.

As in cuprates\cite{devereaux96}, there are two bands in the bilayer one-orbital model, 
\begin{align}
e_{1,2}({\bf k})= -2 t (\cos k_x + \cos k_y) \pm t_\perp- \mu,
\end{align}
\noindent split by $t_\perp$. 

Here, we show the Raman response using the same procedure discussed Sec. \ref{sec:IB}, i.e., the IB approximation which had shown to be appropriate for cuprates\cite{devereaux07}. Since the transformation that diagonalizes the Hamiltonian does not depend on $k_x$ and $k_y$, the IB approximation is reliable.
Although the  present calculation follows the lines of Ref. \cite{devereaux96}, it is instructive to present these results in order to compare with the multiorbital case.

\begin{figure*}[htbp]
    \centering
        \includegraphics[width=16.0 cm]{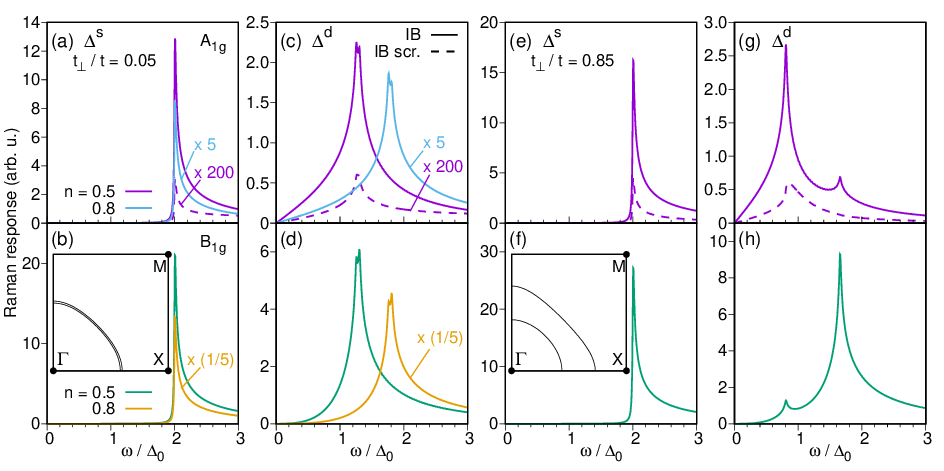}
        \caption{$A_{1g}$ and $B_{1g}$ Raman responses for the bilayer one-orbital model for $s$- and $d$-wave pairing. 
    Solid lines are the results for the free Raman responses, while dashed lines show the screened $A_{1g}$. (a)-(d) are the results for $t_{\perp}/t=0.05$ and fillings $n=0.5$ and $n=0.8$.
    (e)-(h) are for $t_{\perp}/t = 0.85$ and $n=0.5$. Insets in (b) and (f) show the two $n = 0.5$ FS sheets for $t_\perp/t=0.05$ and $t_\perp/t=0.85$, respectively.}
    \label{fig:screening}
\end{figure*}

In the following, and for qualitative and semiquantitative comparisons, we consider a gap $\Delta_0/t=0.05$ which can be thought of the order of the superconducting gap in nickelates. For instance, using an effective hopping parameter $t = 0.25$ eV, smaller than the bare one\cite{gu23,luo23}, reflecting the effective mass renormalization due to correlation, the value of $\Delta_0$ is of the order to be consistent with $T_c = 80$ K.
We discuss two different gaps, an $s$-wave gap $\Delta^s=\Delta_0$ and a $d$-wave gap $\Delta^d=\Delta_0 \gamma_d({\bf k})$ for each band, where $\gamma_d({\bf k})=(1/2)(\cos k_x - \cos k_y)$. Since we did not include a second-nearest-neighbors hopping $t'$, in present paper we do not consider the $B_{2g}$ Raman response, which is expected to show a much weaker contribution than $A_{1g}$ and $B_{1g}$ symmetries because the  $B_{2g}$ is proportional to $t'^2$ and $t' \ll t$.

In Fig. \ref{fig:screening}(a)-\ref{fig:screening}(d) we consider two fillings and $t_\perp/t = 0.05$, which is similar to the small interlayer hopping between $d_{x^2-y^2}$ orbitals of different planes for bilayer nickelates \cite{lu24,qu24,oh24t,bejas25}, and of the order of that in infinite-layer nickelates\cite{botana20}.
The two fillings give, $\mu/t=-1.45$, which corresponds to quarter filling ($n=0.5$ per plane) for the $d_{x^2-y^2}$ orbitals in bilayer nickelates, and $\mu/t=-0.44$, which corresponds to the electron filling $n=0.80$ per plane for infinite-layer nickelates. Temperature is set to zero.
That leads to two Fermi surface (FS) sheets centered at $\Gamma$-point [see inset in Fig. \ref{fig:screening}(b)]. The close proximity between the two FS sheets arise from the small $t_\perp$, which leads to two very close peaks in Fig. \ref{fig:screening}(c)-\ref{fig:screening}(d). 
For both, $A_{1g}$ and $B_{1g}$, the Raman response shows well defined peaks at $2\Delta_0$ for $s$-wave pairing.
While for $n=0.8$ the intensity of the $A_{1g}$ Raman response is much lower than for $n=0.5$, for $B_{1g}$ it is the opposite.
For the $d$-wave case, two peaks at $\omega \lesssim 2\Delta_0$ arise from the value of the gap at the Fermi surface of each band. The peaks are very close together due to the small splitting.
Similar to the $s$-wave gap, for $n=0.8$ the intensity of the $A_{1g}$ Raman response is much lower than for $n=0.5$ and it is the opposite  for $B_{1g}$.
In addition, for $n=0.8$ there is a tendency to form a peak at higher energy than for $n=0.5$. 
A linear and cubic power law at low energy is observed for $A_{1g}$ and $B_{1g}$, respectively.
Importantly, the screened $A_{1g}$ response [dashed lines in Figs. \ref{fig:screening}(a) and \ref{fig:screening}(c)] is negligible for both, $s$-wave and $d$-wave. 
Here we have shown only results for $n=0.5$ because for $n=0.8$ the $A_{1g}$ response is similarly screened.
Thus, $A_{1g}$ signal is not expected to be observed in both cases, and a stronger $B_{1g}$ peak is predicted for infinite-layer nickelates than for bilayer nickelates.

We also consider the bilayer model with a strong interlayer coupling, which can be realized in systems with active $d_{z^2}$ orbitals. This bilayer Hubbard model exhibits competing $d$-wave and $s_{\pm}$-wave superconductivity away from half-filling~\cite{PhysRevB.77.144527,PhysRevB.75.193103,PhysRevB.80.064517,PhysRevB.84.180513}. Our Raman results with a strong interlayer hopping $t_{\perp}/t=0.85$ for $s$-wave and $d$-wave pairings are displayed in Figs. \ref{fig:screening}(e)-\ref{fig:screening}(h) for comparison with the $t_{\perp}/t=0.05$ case.
We chose $\mu/t=-1.53$, which correspond to quarter filling $n=0.5$. 
The results are somewhat similar to the results for $t_{\perp}/t=0.05$. For instance, for an $s$-wave gap a sharp peak is obtained for both, $A_{1g}$ and $B_{1g}$ at $2\Delta_0$.   
However there are two main differences.
(i) For both, $A_{1g}$ and $B_{1g}$ channels for the $d$-wave symmetry gap [Figs. \ref{fig:screening}(g) and \ref{fig:screening}(h)] it is possible to see two well separated peaks, each one corresponding to a given band, due to the large $t_{\perp}$. In the inset of Fig. \ref{fig:screening}(f) it is possible to see the two FS sheets for $t_\perp/t=0.85$.
While the lowest-intensity peak in $A_{1g}$ occurs at the same energy of the highest-energy peak in $B_{1g}$, the opposite is obtained for the highest-intensity peak in $A_{1g}$.   
From an experimental point of view, it is expected that the low-intensity peaks will not be visible in the experiments because they can be easily blurred out by experimental resolution, or other intrinsic effects that can be simulated by a larger broadening $\Gamma$.
(ii) The screening in the $A_{1g}$ channel [see dashed lines in Figs. \ref{fig:screening}(e) and \ref{fig:screening}(g)] is much less efficient than for $t_{\perp}/t=0.05$, allowing a larger Raman response. For instance, while for $t_{\perp}/t=0.05$ the screened Raman response is more than $\sim 200$ times weaker that the free $A_{1g}$ response, for $t_{\perp}/t=0.85$ the screened response is only about $\sim 4$ times lower than the free one. The reason for the  more efficient screening for $t_{\perp}/t=0.05$ than for $t_{\perp}/t=0.85$ is the following. For the case of a free electron gas (dispersion $k^2/2m$) the $A_{1g}$ screened response is just zero, i.e., the second term in Eq. (\ref{Rasauer}) cancels exactly the first term (Ref. \cite{sauer82}).
For small splitting $t_\perp/t=0.05$ we have two quasicircular FS sheets centered at $\Gamma$ [see inset in Fig. \ref{fig:screening}(b)] which resembles the spherical FS of the free electron gas.
Instead, for $t_\perp/t=0.85$ the two FS sheets look further away
from the free electron gas [see inset in Fig. \ref{fig:screening}(f)] and the screening is less efficient than for $t_\perp/t=0.05$.

\section{Single-layer two-orbital model}\label{sec:single2orb}
In this section we study the model which was recently proposed for discussing the impact of the charge density wave on Raman experiments in nickelates\cite{CDW}.
The proposed model, which can be considered as a simplified model for one-layer of NiO$_2$, was defined in Sec. \ref{sec:fullmulti} assuming that $x$ ($z$) represents the orbital Ni-$d_{x^2-y^2}$ (Ni-$d_{z^2}$). This model is also named as the $m=1$ case in Ref. \cite{zhang25DD}.

\begin{figure}[htbp]
   \centering
   \includegraphics[width=8.6 cm]{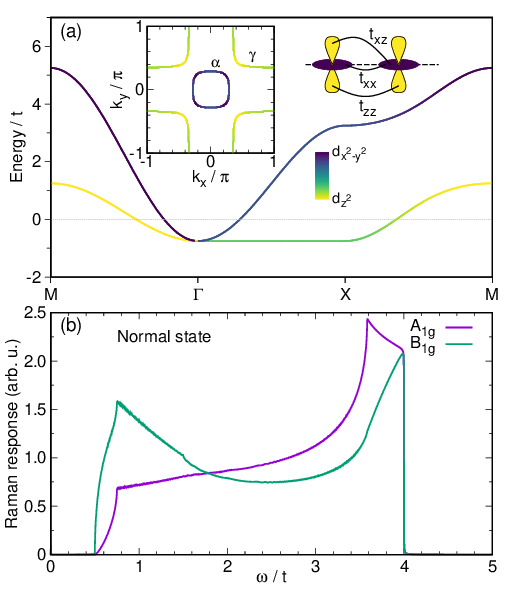}
   \caption{(a) Energy bands in the single-layer two-orbital model. Left inset shows the Fermi surface sheets, where the orbital contributions  are represented by colors. The right inset sketches the model.
   (b) Raman response in the $A_{1g}$ and $B_{1g}$ channels in the normal state. }
   \label{fig:FSynormal}
\end{figure}

The explicit form for the hopping terms in this case are:
$t_{xx}({\bf k})= -(3t/2) (\cos k_x + \cos k_y) + \epsilon - \mu$ is the hopping between Ni-$d_{x^2-y^2}$ orbitals, 
$t_{zz}({\bf k}) =-(t/2) (\cos k_x + \cos k_y) - \epsilon - \mu$ is the hopping between Ni-$d_{z^2}$ orbitals, and
$t_{xz}({\bf k}) = (\sqrt{3}t/2) (\cos k_x - \cos k_y)$ is the hopping between Ni-$d_{x^2-y^2}$ and Ni-$d_{z^2}$ orbitals.
$\epsilon$ is the orbital splitting and $\mu$ the chemical potential. 
${\bf k}$ is the momentum in the 2D square lattice where on each site the Ni-$d_{x^2-y^2}$ and Ni-$d_{z^2}$ orbitals are located.
The right inset in Fig. \ref{fig:FSynormal}(a) sketches the model.   

The Hamiltonian in the band basis $H_{\bf k}^B$ is diagonal and  $e_1 = (t_{xx} + t_{zz})/2 + r$ and $e_2 = (t_{xx} + t_{zz})/2 - r$ are the two energy bands, where $r = \sqrt{(t_{xx}-t_{zz})^2/4
+ t_{xz}^2}$. The momentum ${\bf k}$ was omitted for convenience.
In the following, we fix the same parameters as those used in Ref. \cite{CDW}. The orbital splitting is $\epsilon/t= 1$ and the electron density $n=4/3$, which corresponds to $\mu/t = -1.25$.
In Fig. \ref{fig:FSynormal}(a) we present the bands, and the Fermi surface (FS) which represents qualitatively the $\alpha$ and $\gamma$ sheets of the FS in nickelates\cite{gu23}. The orbital contributions  are represented by colors.
In Fig. \ref{fig:FSynormal}(b) we show the Raman response for $A_{1g}$ (purple line) and $B_{1g}$ (green line) in the normal state, i.e, the superconducting gaps are zero. Only interband transitions play a role, i.e., those that contain $R^B_{12}$ and $R^B_{21}$. As can be seen, interband transitions occur above an onset energy 
$\omega/t \sim 0.5$. 

In the superconducting state low-energy pair breaking features are expected at low energy below the onset of interband transitions.  As we are interested in the low-energy pair breaking features we present results up to $\omega=3\Delta_0$. 
\begin{figure*}[htbp]
    \centering
    \includegraphics[width=16 cm]{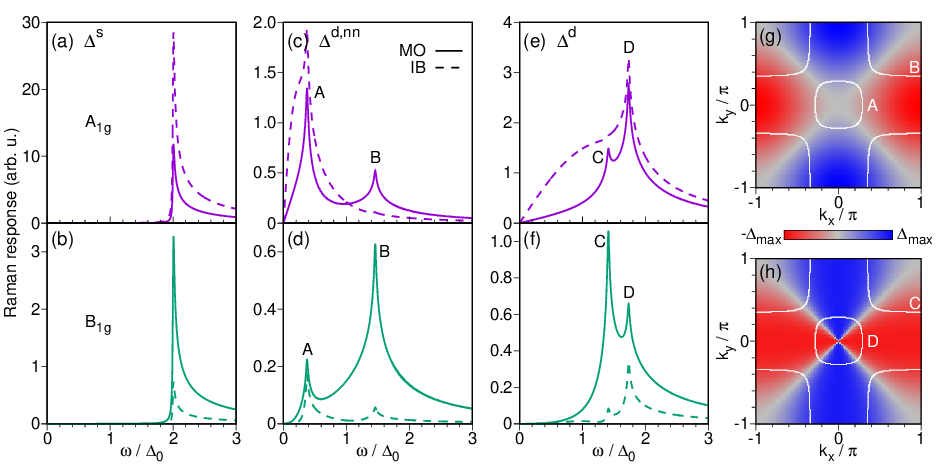}
    \caption{Solid lines are $A_{1g}$ [(a), (c), (e)] and $B_{1g}$ [(b), (d), (f)] Raman responses for different pairings. Dashed lines show the results for the additive Raman response where each band is considered separately.
    (a) and (b) intraorbital $s$-wave pairing,
    (c) and (d) intraorbital $d$-wave pairing on nearest-neighbors bonds,
    (e) and (f) interorbital $d$-wave pairing.
    (g) and  (h) Comparison of the intraband pairing for the intraorbital $d$-wave with nearest-neighbors bonds (g) and interorbital $d$-wave (h).
    White lines represent the FS from the inset in Fig. \ref{fig:FSynormal}(a).
    Labels A-D indicate the origin of the peaks in panels (c)-(f).}
    \label{fig:results}
\end{figure*}
Next, we present results for the following gaps: 
\begin{align}
\Delta^s = \Delta_0 \tau_0 , \quad
\Delta^{d,nn} = \Delta_0 \gamma_d({\bf k}) \tau_0, \quad
\Delta^d = \Delta_0 \tau_1 \, ,
  \label{eq:gapsintraorb}
\end{align}
\noindent where $\tau_0$ is the $2\times 2$ Pauli identity matrix. In Eq. (\ref{eq:gapsintraorb}) the Pauli matrices act on the orbital basis. Since our aim is to discuss qualitative features we do not considered a possible gap mixing, which can be easily included if necessary.

For $\Delta^s$ (isotropic intraorbital s-wave pairing), the $A_{1g}$ channel [solid purple line in
Fig. \ref{fig:results}(a)]
shows a sharp pair breaking peak at twice the superconducting gap $2\Delta_0$. For the $B_{1g}$ channel [solid green line in
Fig. \ref{fig:results}(b)] 
the pair breaking peak is also located at $2\Delta_0$, although the Raman intensity is lower than for $A_{1g}$.

For $\Delta^{d,nn}$ (intraorbital $d$-wave pairing on nearest-neighbors bonds), interestingly, two low-energy peaks are observed in $A_{1g}$ [solid purple line in Fig. \ref{fig:results}(c)] and $B_{1g}$ [solid green line in Fig. \ref{fig:results}(d)].
In both symmetries the lower peak is located at energy $\omega \sim \Delta_0/2$ and the upper one at $\omega \sim 1.5 \Delta_0$. The low-energy peak is related to the $\alpha$ sheet of the FS and the high-energy one to the $\gamma$ sheet. 

For $\Delta^d$ (interorbital $d$-wave pairing) there is still a pair breaking feature at $\sim 1.5\Delta_0$ for $A_{1g}$
[solid purple line in Fig. \ref{fig:results}(e)] and $B_{1g}$ [solid green line in Fig. \ref{fig:results}(f)].
The low-energy feature for the intraorbital $d$-wave pairing on nearest-neighbors bonds (point A) occurs now at higher energy (point D).
In Fig. \ref{fig:results}(g) and Fig. \ref{fig:results}(h) we show the momentum dependence of the gap in each case. Although both show the characteristic $d$-wave structure and their values close to the border of the Brillouin zone are similar, the intraorbital $d$-wave pairing on nearest-neighbors bonds gap takes lower values close to the $\Gamma$ point [Fig. \ref{fig:results}(g)].
Following this, the peak from the $\gamma$ sheet of the FS is located at roughly the same energy in both cases (points B and C), while the peak related to the $\alpha$ sheet appears at low energy in the $\Delta^{d,nn}$ case (point A) and at high energy in the $\Delta^d$ case (point D).
Interestingly, both $d$-wave Raman responses follow the same power law at low energy, i.e., $\sim \omega$ and $\sim \omega^3$ for $A_{1g}$ and $B_{1g}$, respectively. 
For the case of the one-band model it is well known that $d$-wave symmetry exhibits the power laws $\sim \omega$ and $\sim \omega^3$ for $A_{1g}$ and $B_{1g}$, respectively (see Refs. \onlinecite{devereaux07,andreasthesis}, where this is discussed extensively). Here we have shown that the same occurs for the two-orbital case. In this aspect the IB calculation is very useful. Since in the IB calculation the Raman response is calculated band-by-band separately, each one fulfills the analysis of Refs. \onlinecite{devereaux07,andreasthesis}. Then, although the analytical probe in multiorbital systems is difficult, combining numerical calculations and the IB response we suggest that the power laws follow the same characteristics as for the one-orbital model.
For $A_{1g}$, in contrast to what it is expected in single-layer and single-orbital cases\cite{devereaux96}, the Coulomb screening, the second term in Eq. (\ref{Rasauer}), is not efficient giving a negligible contribution. 

For intraorbital $s$-wave pairing ($\Delta^s$) [Fig. \ref{fig:results}(a), \ref{fig:results}(b)] both approximations, the MO and the IB, give similar results. Although with different intensity, they show peaks at $\sim 2\Delta_0$ in both channels, $A_{1g}$ and $B_{1g}$. 
For the intraorbital $d$-wave pairing on nearest-neighbors bonds [Fig.\ref{fig:results}(c), \ref{fig:results}(d)] the results from both approximations are qualitatively different. While both approximations show a low peak at $\omega \sim \Delta_0/2$, the second peak at about $1.5\Delta_0$ is suppressed in the IB approach.
In addition, the IB approach predicts a broader low-energy peak for $A_{1g}$ than the MO method.
For the interorbital $d$-wave pairing [Fig.\ref{fig:results}(e), \ref{fig:results}(f)] the results from both approximations show
similar differences, with the $\gamma$ sheet peak (B and C) strongly suppressed in the IB approximation.
For $A_{1g}$ a larger spectral weight is predicted from the IB than from  MO method at low energy. In addition, the Raman response in the  $B_{1g}$ channel is much weaker in the IB than in the MO method. 

In the particular case for the B$_{1g}$ spectra for the intraorbital $d$-wave pairing on nearest-neighbors bonds [Fig.\ref{fig:results}(d)], the difference between the MO and the IB methods is more striking, and deserves some comments. 
The MO calculation predicts two peaks, and in principle the most promising candidate to be detected in the experiment is the more intense one at $\omega \sim 1.5\Delta_0$ (point B), however in the IB approximation this peak is heavily suppressed.
As the intraband transitions in both calculations are the same, the difference in the peak intensity comes then from the Raman vertices computed by each method.
Care should be taken when using the IB approximation in multiorbital systems because some features can be qualitatively different respect to the MO calculation.

Finally, it is important to mention that the discussed differences between the MO and the IB approximations 
also occurs for other superconducting gap values. We have also checked these pictures (not shown) for $\Delta_0/t=0.1$, and $\Delta_0/t=0.01$.

\section{Bilayer two-orbital model}\label{sec:bi2orb}

\begin{figure}[htbp]
   \centering
   \includegraphics[width=8.6 cm]{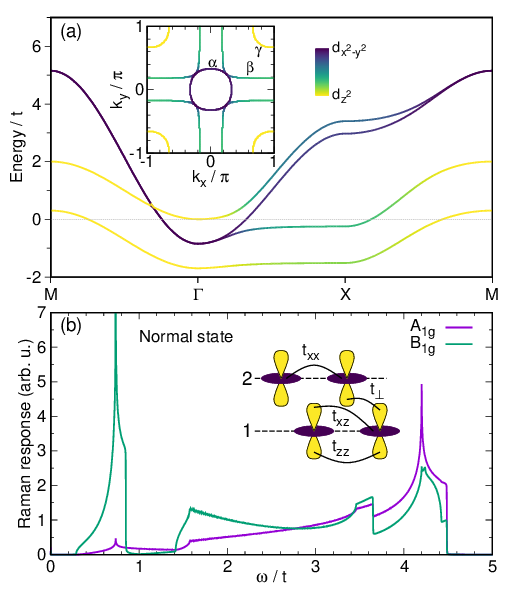}
   \caption{(a) Energy bands in the bilayer two-orbital model. The inset shows the Fermi surface sheets. The orbital contributions are represented by colors.
   (b) Raman response in the $A_{1g}$ and $B_{1g}$ channels in the normal state. The inset sketches the model.}
   \label{fig:FSynormalBi}
\end{figure}

The proposed model Hamiltonian is given by a bilayer formed by the stacked of two single-layer two-orbital models discussed in Sec. \ref{sec:single2orb}, and it is written as 
$H_{0} = \sum_{{\bf k},l}  \psi_{{\bf k},l}^\dag H_{{\bf k},l} \psi_{{\bf k},l} + \sum_{{\bf k}} H^\perp_{\bf k} $.  In $H_{0}$ the index $l$ takes the values $1$ and $2$ corresponding to the two layers. 
$H_{{\bf k},l}$ has the same form as Eq. (\ref{eq:Hk}) for each layer, and $\psi_{{\bf k},l}^\dag = (x_{{\bf k},l}^\dag, z_{{\bf k},l}^\dag)$.
$H^\perp_{\bf k}= t_\perp (z_{{\bf k},1}^\dag z_{{\bf k},2} + {\rm H.c})$, where $t_\perp$ is the interlayer hopping between $d_{z^2}$ orbitals, which is well known to be large and even of the order of $t$\cite{luo23,gu23}. Thus, this model is just a simple and natural extension of the model of Sec. IV to two layers coupled with the larger interlayer hopping between $d_{z^2}$ orbitals. In the following, we chose $t_\perp/t=-0.85$ and $n = 8/3$ which corresponds to $\mu/t = -1.16$. The inset in Fig. \ref{fig:FSynormalBi}(b) sketches the model. Although this model seems to be the simpler bilayer one, it was extensively discussed in the context of nickelates\cite{zhang24,maier25}, and called $m=2$ in Ref. \cite{zhang25DD}. In particular, in Ref. \cite{zhang24} the model was studied in the context of superconductivity. Then, the results of the present section can also be considered as Raman predictions for this study. Figure \ref{fig:FSynormalBi}(a) shows the four energy bands. The inset shows three FS sheets which are very similar to the $\alpha$, $\beta$, and $\gamma$ sheets discussed and observed in nickelates\cite{zhang24,maier25,gu23,Wang25RR}. The orbital contributions are represented by colors.
\begin{figure*}[tbp]
    \centering
    \includegraphics[width=16 cm]{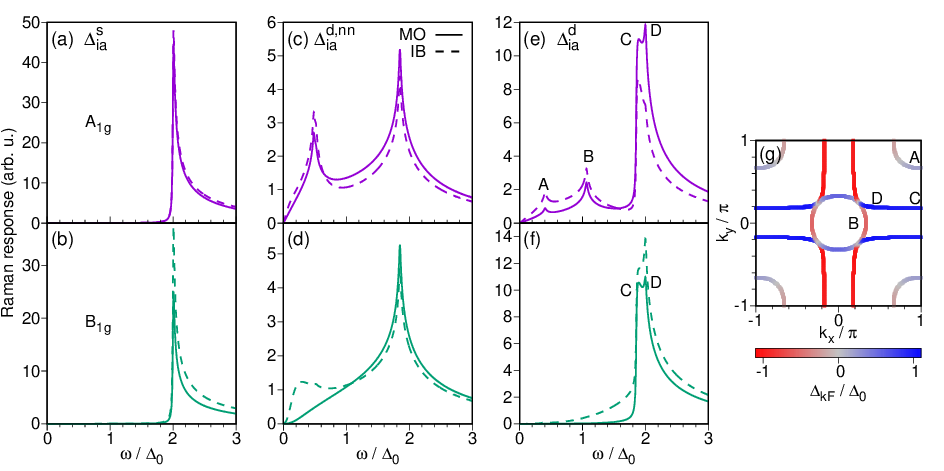}
    \caption{Solid lines are $A_{1g}$ [(a), (c), (e)] and $B_{1g}$ [(b), (d), (f)] Raman responses for different pairing symmetries. Dashed lines show the results for the additive Raman response where each band is considered separately.
    (a) and (b) intralayer intraorbital $s$-wave pairing,
    (e) and (f) intralayer interorbital $d$-wave pairing.
    (g) gap value along the FS for the paring in (e) and (f). Labels A-D indicate the region of the FS that produces the different peaks in (e), while C and D are only observed in (f).}
    \label{fig:resultsBiintra}
\end{figure*}
The procedure for calculating the Raman response follows the same formalism as discussed in Sec. \ref{sec:fullmulti} but now the Hamiltonian, pairing, and Raman vertices in the orbital and band space are $4 \times 4$ matrices, while the Nambu Green's function is an $8 \times 8$ matrix.

In Fig. \ref{fig:FSynormalBi}(b) we show the Raman response for $A_{1g}$ (purple line) and $B_{1g}$ (green line) in the normal state, i.e, the superconducting gaps are zero. Only interband transitions contribute, and occur above an onset energy $\omega/t \sim 0.3$ in both channels.

\begin{figure}[htbp]
    \centering
    \includegraphics[width=8.6 cm]{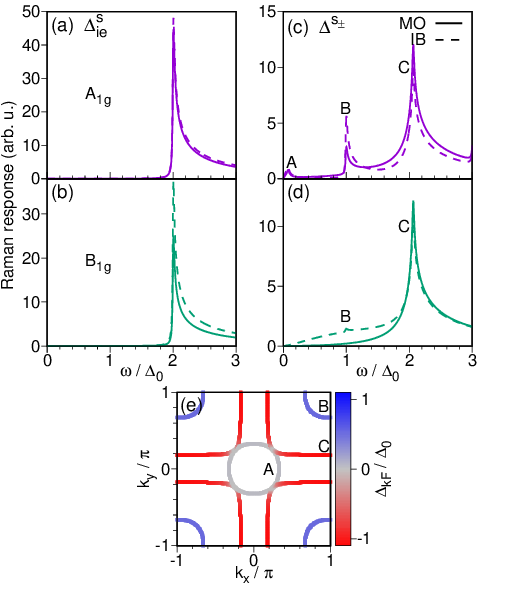}
    \caption{Solid lines are $A_{1g}$ [(a), (c)] and $B_{1g}$ [(b), (d)] Raman responses for different pairings. Dashed lines show the results for the additive Raman response where each band is considered separately.
    (a) and (b) interlayer intraorbital $s$-wave pairing,
    (c) and (d) interlayer-dominated $s^{\pm}$-wave pairing on the $d_{z^2}$ orbital.
    (e) gap value along the FS for the paring in (c) and (d). Labels A-C indicate the region of the FS that produces the different peaks in (c) and (d).}
    \label{fig:resultsBiinter}
\end{figure}

Similar to the single-layer two-orbital model, and for illustration, we consider several and basic candidates for pairing states, classified according to their layer, orbital structure,  and symmetry. The corresponding superconducting order parameters are given by:
\begin{align}
\Delta_{\textrm{ia}}^s &=
    \Delta_0 \tau_0 \otimes \tau_0,\quad
\Delta_{\textrm{ia}}^{d,nn}= 
    \Delta_0 \gamma_\mathbf{k} \tau_0 \otimes \tau_0,\quad  \nonumber \\
\Delta_{\textrm{ia}}^d &= 
    \Delta_0 \tau_0 \otimes \tau_1,\quad
\Delta_{\textrm{ie}}^s= 
    \Delta_0 \tau_1 \otimes \tau_0,\quad  \nonumber \\
\Delta_{\textrm{ie}}^d &= 
    \Delta_0 \tau_1 \otimes \tau_1,\quad  \nonumber \\
\Delta^{s_{\pm}} &= 
		\tau_0
		\otimes
		\begin{pmatrix}
			0 & 0\\
		    0 & -\Delta_0/2
		\end{pmatrix}
        +
        \tau_1
		\otimes
		\begin{pmatrix}
			0 & 0\\
		      0 & \Delta_0
		\end{pmatrix} \, .
 \label{eq:gapsbitwo}
\end{align}
In Figs. \ref{fig:resultsBiintra} and \ref{fig:resultsBiinter} we present the Raman responses at low energy where pair breaking features are expected. While in Fig. \ref{fig:resultsBiintra} we present results for different intralayer pairings (ia),
in Fig. \ref{fig:resultsBiinter} we show results for the interlayer (ie) intraorbital $s$-wave pairing $\Delta_{\textrm{ie}}^s$ and $\Delta^{s_{\pm}}$.

For $\Delta_{\textrm{ia}}^s$ [intralayer intraorbital $s$-wave pairing, Fig. \ref{fig:resultsBiintra}(a) and \ref{fig:resultsBiintra}(b)] and $\Delta_{\textrm{ie}}^s$ [interlayer intraorbital $s$-wave pairing, Fig. \ref{fig:resultsBiinter}(a) and \ref{fig:resultsBiinter}(b)], both $A_{1g}$ and $B_{1g}$ show similar sharp peaks at $2\Delta_0$. In addition, these results are also close to the case for single-layer [Fig. \ref{fig:results}(a) and \ref{fig:results}(b)]. Thus, it seems difficult to distinguish between these cases from the experiment.

For $\Delta_{\textrm{ia}}^{d,nn}$ (intralayer intraorbital $d$-wave on nearest-neighbors bonds) and the $A_{1g}$ channel, similar to the single-layer case [Fig. \ref{fig:results}(c)], Fig. \ref{fig:resultsBiintra}(c) shows a low-energy peak at about $\sim \Delta_0/2$ with a linear law at low energy, and a second peak at higher energy $\sim 2 \Delta_0$. However, opposite to the single-layer case, the intensity of this second peak is larger than the first peak.
For $B_{1g}$, both the bilayer [Fig. \ref{fig:resultsBiintra}(d)] and the single-layer [Fig. \ref{fig:results}(d)] model show a large peak at about $\sim 2\Delta_0$.

For $\Delta_{\textrm{ia}}^{d}$ (intralayer interorbital $d$-wave pairing) Fig. \ref{fig:resultsBiintra}(e) and \ref{fig:resultsBiintra}(f) show sharp peaks at $\sim 2\Delta_0$ (peaks C and D), although for $A_{1g}$ a sizable spectral weight in the form of peaks is observed at low energy (peaks A and B), which is missing in $B_{1g}$.
In addition, for the single-layer case a $\sim \omega$ ($\sim \omega^3$) power law is more clearly obtained for $A_{1g}$ ($B_{1g}$) than for the bilayer.
In Fig. \ref{fig:resultsBiintra}(g) we show the gap values on each FS sheet.
The peaks denoted by A, B, C, and D in Fig. \ref{fig:resultsBiintra}(e), and C and D in Fig. \ref{fig:resultsBiintra}(f) come from the corresponding ones in Fig. \ref{fig:resultsBiintra}(g).
The absence of peaks A and B in the $B_{1g}$ channel is due to the momentum dependence of the $B_{1g}$ Raman vertex, yielding low spectral weight in the momentum region from which these peaks originate.
Very similar results (not shown) are obtained for $\Delta_{\textrm{ie}}^{d}$ (interlayer interorbital $d$-wave pairing).

Figures \ref{fig:resultsBiinter}(c) and \ref{fig:resultsBiinter}(d) show results for $\Delta^{s_\pm}$ (interlayer-dominated $s^{\pm}$-wave pairing on the $d_{z^2}$ orbital\cite{zhan25pp}) for $A_{1g}$ and $B_{1g}$, respectively. 
Similar to the $\Delta^d_{\rm ia}$ case, both channels show a sharp peak at $\sim 2\Delta_0$ (peak C), and $A_{1g}$ shows extra peaks at low energy (peaks A and B) not visible in the $B_{1g}$ spectra due to the corresponding Raman vertex.
A tiny remnant of the peak B appears in $B_{1g}$ in the IB approximation.
In  Fig. \ref{fig:resultsBiinter}(e) we show the gap values on each FS sheet for $s^{\pm}$.
The peaks denoted by A, B, and C in Fig. \ref{fig:resultsBiinter}(c) and \ref{fig:resultsBiinter}(d) come from the corresponding labeled sectors of the FS in Fig. \ref{fig:resultsBiinter}(e). Despite the simplicity of our bilayer model, it is interesting to make comparisons with results obtained in the framework of more complicated models. Recently\cite{zhanjun25}, the pair breaking features were discussed for a more complicated model, with more hopping parameters between orbitals obtained by DFT calculations\cite{gu23}. Comparing our Figs. \ref{fig:resultsBiinter}(c) and \ref{fig:resultsBiinter}(d) with Fig. 3 in Ref. \cite{zhanjun25}, we can see that the results share some common features. In both cases the leading peak is at $2\Delta_0$ and some subleading peaks are expected in the energy scale of $\Delta_0$.
A similar analysis as in Figs. \ref{fig:resultsBiintra}(g) and \ref{fig:resultsBiinter}(e) can be performed for all the gap symmetries in different channels.

We have also presented results for the IB approximation for each case (dashed lines). Interestingly, the IB reproduces better the results for the MO calculation than for the single-layer two-orbital case.
For completeness, we performed calculations (not shown) for a small $t_\perp/t=-0.05$ and the results look very similar to all cases discussed for the single-layer two-orbital model (Sec. \ref{sec:single2orb}) showing the importance of a large $t_\perp$ for the bilayer two-orbital model.
\vspace{1 cm}

\section{Conclusion}\label{sec:conc}
Mainly motivated by the recent discovery of superconductivity at high critical temperature $T_c$ in pressurized and thin film nickelates, we have discussed the Raman response in the superconducting phase for different pairings and simple single-layer and bilayer two orbital models, and a bilayer single-orbital model.
Although we have studied some basic pairing symmetries ($s$-wave, $d$-wave, and $s^{\pm}$-wave), our study shows features in the Raman spectra that can be used for discussing the gap size and its symmetry, and the possible minimal model for superconductivity.
Given the recent discovery of high-$T_c$ superconductivity in bilayer nickelates at intermediate pressure and in thin films nickelates at ambient pressure, Raman experiments across $T_c$ are expected in the near future.
Our generic calculation can  be  adapted easily to these studies and different more complicated proposed models.
Thus, the  method is potentially useful for analyzing experiments, providing information of great interest for understanding the mechanisms that lead to high-$T_c$ superconductivity in nickelates. For instance, if Raman experiments are available we can contribute to solve the controversy presented in the recent reports\cite{guo25,cao25}, which discuss different superconducting gap symmetries and values.
We performed our analysis using two different approaches, the full multiorbital calculations and the additive Raman response where each band is considered separately.
We concluded that in some cases, the results obtained by using the additive Raman response approximation should be taken with caution due to the differences  arising from the different method of computing the Raman vertices.
Finally, our discussions are also of interest for other multiorbital systems distinct to nickelates. In addition, Higgs mode resonances in optical response of one-band and multiband superconductors have recently attracted
attention, see, for instance, Refs. \cite{kim24,kaj23,fiore22}. Since nickelates are multiband high-$T_c$ superconductors and possess a large gap, these materials may offer a good platform for these studies.

\begin{acknowledgments}
We thank M. Hepting, G. Khaliullin, and V. Sundaramurthy for useful discussions.
Parts of the results presented in this work were obtained by using the facilities of the CCT-Rosario Computational Center, member of the High Performance Computing National System (SNCAD, MincyT-Argentina). A.G., M.B., J.Z., and X.W.  acknowledge the Max-Planck-Institute for Solid State Research in Stuttgart for hospitality and financial support. 
A.P.S. is funded by the Deutsche Forschungsgemeinschaft (DFG, German Research Foundation) – TRR 360 – 492547816. X.W. is supported by the National Key R\&D Program of China (Grant No. 2023YFA1407300) and the National Natural Science Foundation of China (Grants No. 12574151, 12447103 and 12447101). 
\end{acknowledgments}


\bibliography{main_longpaper}

\end{document}